\documentclass[aps, prb, 10pt, superscriptaddress, twocolumn]{revtex4-1}

\usepackage{amsmath,amssymb}
\usepackage{graphicx}

\DeclareMathOperator{\re}{Re}
\DeclareMathOperator{\im}{Im}

\allowdisplaybreaks[1]

\begin{document}

\title{Fisher--Hartwig expansion for Toeplitz determinants and the spectrum
of a single-particle reduced density matrix for one-dimensional free fermions}

\author{Dmitri A.~Ivanov}
\affiliation{Institute for Theoretical Physics, ETH Zurich, 8093 Zurich, Switzerland}
\affiliation{Institute for Theoretical Physics, University of Zurich, 8057 Zurich, Switzerland}

\author{Alexander G.~Abanov}
\affiliation{Department of Physics and Astronomy and Simons Center for Geometry and Physics,
Stony Brook University,  Stony Brook, NY 11794, USA}

\date{June 20, 2013}

\begin{abstract}
We study the spectrum of the Toeplitz matrix with a sine kernel, which corresponds
to the single-particle reduced density matrix for free fermions on the one-dimensional lattice.
For the spectral determinant of this matrix, a Fisher--Hartwig expansion in the inverse
matrix size has been recently conjectured. This expansion can be verified order by order,
away from the line of accumulation of zeros, using the recurrence relation known from
the theory of discrete Painlev\'e equations.
We perform such a verification to the tenth order and calculate the corresponding
coefficients in the Fisher--Hartwig expansion. 
Under the assumption of the validity 
of the Fisher--Hartwig expansion in the whole range of the spectral parameter,
we further derive expansions for an equation on the eigenvalues of this matrix and 
for the von Neumann entanglement entropy
in the corresponding fermion problem.
These analytical results are supported by a numerical example.
\end{abstract}



\maketitle

\section{Introduction}

Toeplitz matrices, i.e., matrices with entries $a_{i-j}$ depending only on the difference
of the column and row indices, play an important role in a wide range of physical and
mathematical problems, including problems in statistical mechanics \cite{montroll:63,mccoywu:67,basor:06},
random-matrix theory \cite{tracy:93,widom:94}, quantum integrable systems 
\cite{its:90.1,its:90.2,its:92,its:93,deift:97,goehmann:98,fujii:00,cheianov:04}, and 
nonequilibrium bosonization \cite{gutman:10,gutman:11,protopopov:12}.
One is often interested in the asymptotic behavior of the
spectrum of such matrices in the limit of a large matrix size $L$, for a given
{\em symbol} of the matrix
\begin{equation}
\sigma(k) = \sum_{m=-\infty}^{+\infty} a_m e^{-ikm}\, , 
\qquad k\in [-\pi,\pi]\, .
\label{symbol-general}
\end{equation}
While the leading exponential dependence of the spectral determinant
of Toeplitz matrices on their size $L$ is usually easy to guess on physical grounds
(the rigorous result is known as the first Szeg\H{o} theorem) \cite{szego:15}, finding 
subleading corrections is a nontrivial problem, which continues to be a topic of active research
in mathematics and mathematical physics \cite{szego:52,fisher:68,basor:78,basor:83,basor:91,%
boettcher:94,ehrhardt:01,boettcher:06,krasovsky:11}. A recent important step
in developing the theory of Toeplitz determinants was the
proof of the Fisher--Hartwig conjecture for the case of symbols
(\ref{symbol-general}) with power-law singularities \cite{deift:11}.

In mathematical and physical literature, stronger conjectures about higher-order
corrections to the Fisher--Hartwig formula have been proposed 
\cite{abanov:03:franchini:05,calabrese:10,gutman:11,kozlowski:08,kitanine:09.1,kitanine:09.2,kozlowski:10,abanov:11}. 
In particular, many studies focused on the specific example of the matrix of the {\em sine-kernel}
form:
\begin{equation}
a_{i-j}= \begin{cases}
\frac{\sin k_F(i-j)}{\pi(i-j)}\, , & i\ne j\, , \\
k_F/\pi \, , & i=j\, ,
\end{cases}
\label{matrix-a}
\end{equation}
which corresponds to the symbol
\begin{equation}
\sigma(k)=
\begin{cases}
1\, , & |k| < k_F\, , \\
0\, , & {\rm otherwise}\, .
\end{cases}
\end{equation}

This Toeplitz matrix (with $k_F$ being the only parameter,
apart from the matrix size $L$) appears in many problems
involving one-dimensional free fermions and closely related
systems. The parameter $k_F$ corresponds to the Fermi
wave vector of the fermions. In Ref.~\onlinecite{abanov:11}, 
the higher-order corrections to the spectral determinant of this matrix,
\begin{equation}
\chi(\kappa,k_F,L)=\det_{1 \le i,j \le L} [\delta_{ij}+ a_{i-j} (e^{2\pi i \kappa}-1)]\, ,
\label{spectral-determinant}
\end{equation}
were conjectured to form an asymptotic series
\begin{equation}
\chi(\kappa,k_F,L)=\sum_{j=-\infty}^{+\infty} \chi_* (\kappa+j, k_F, L)\,
\label{expansion-1}
\end{equation}
where
\begin{multline}
\chi_* (\kappa, k_F, L) = \exp \Big[
2 i \kappa k_F L - 2 \kappa^2 \ln (2L\sin k_F) \\
 + {\tilde C}(\kappa)
+\sum_{n=1}^{\infty} {\tilde F}_n(\kappa,k_F)\, (i L)^{-n} \Big]\, ,
\label{expansion-2}
\end{multline}
\begin{equation}
{\tilde C}(\kappa) = 2 \ln [G(1+\kappa) G(1-\kappa)]\, ,
\end{equation}
and $G()$ is the Barnes G function \cite{note-notations,BarnesG}.
We will call the explicitly periodic in $\kappa$ expansion
(\ref{expansion-1})--(\ref{expansion-2}) the {\em Fisher--Hartwig expansion}.

As pointed out in Ref.~\onlinecite{calabrese:10}, a calculation of
the expansion coefficients ${\tilde F}_n(\kappa,k_F)$ is
possible using a recurrence relation from the
theory of discrete Painlev\'e equations \cite{forrester:03}.
In Ref.~\onlinecite{calabrese:10} such a computation has been
performed to the second order. In the present work, we extend
this calculation to the tenth order and find the coefficients
up to ${\tilde F}_{10}(\kappa,k_F)$. 
Formally, this procedure allows us
to verify, order by order, the periodic form of the expansion 
(\ref{expansion-1})--(\ref{expansion-2}) away from the half-line $\im e^{2\pi i \kappa}=0$,
$\re e^{2\pi i \kappa}<0$. At this half-line, the zeros of $\chi(\kappa)$ accumulate 
(as $L\to\infty$),
which prevents a verification of the expansion at this line, even to a finite order.
However, the available numerical evidence from earlier works \cite{calabrese:10,abanov:11,suesstrunk:12} 
and from this work
(see Section \ref{sec:numerics} below) suggests that the expansion (\ref{expansion-1})--(\ref{expansion-2})
also extends to this half-line. If we assume this conjectured extension, we
can convert the coefficients ${\tilde F}_n(\kappa,k_F)$ into a series expansion
for the eigenvalues of the Toeplitz matrix $a_{i-j}$ 
(or, equivalently, for the zeros of the spectral determinant $\chi(\kappa)$).
Furthermore, under the same assumption, we can derive the power series
for the von Neumann entanglement entropy for free fermions on a segment
of the one-dimensional lattice (thus extending the results of Ref.~\onlinecite{suesstrunk:12}
to the lattice case).

The paper is organized as follows. In the next Section, we summarize the main
results, with references to subsequent sections containing detailed formulas
and derivations.
In Section \ref{sec:fermions}, we overview the relation of the spectral problem for
the Toeplitz matrix to one-dimensional free fermions. In Section \ref{sec:Painleve},
we review the recurrence relation used for calculating the coefficients ${\tilde F}_n(\kappa,k_F)$
and present results of such a calculation to the tenth order.
In Section \ref{sec:eigenvalues}, we use the coefficients ${\tilde F}_n(\kappa,k_F)$
to find the asymptotic expansion of the eigenvalues of the Toeplitz matrix.
In Section \ref{sec:entanglement}, we calculate the expansion for the von Neumann entanglement
entropy. In Section \ref{sec:numerics}, we support
the results of the two preceding Sections
with a numerical example. Finally, the last Section \ref{sec:conclusion} contains several concluding
remarks. 
To avoid lengthy equations, in the main part of the paper we only present the first six orders,
and the results for orders seven to ten are given in the Appendix.

\section{Main results}
\label{sec:main}
The main results of this paper assume the conjectured expansion
(\ref{expansion-1})--(\ref{expansion-2}). Under this assumption,
\begin{itemize}
\item
We compute the coefficients ${\tilde F}_n(\kappa,k_F)$ up to
the tenth ($n=10$) order. They turn out to be polynomials in $\kappa$
and $\cot k_F$. The explicit expressions for ${\tilde F}_n(\kappa,k_F)$
are given in Eqs.\ (\ref{F-coefficients}) and (\ref{p-coefficients})
of Section \ref{sec:Painleve}
and in Eq.~(\ref{p-coefficients-extra}) of the Appendix.
The consistency of this calculation
supports the conjecture (\ref{expansion-1})--(\ref{expansion-2}).
\item
We write an equation on the spectrum of the matrix (\ref{matrix-a})
in the form of a ``quasiclassical expansion'' in $1/L$ 
[Eqs.\ (\ref{spectrum-equation-1}) and (\ref{spectrum-equation-2})]. 
The coefficients
of this expansion are again polynomials in $\kappa$
and $\cot k_F$. The expansion is computed to the tenth order in $1/L$
and the coefficients are reported in Eqs.\ (\ref{X-coefficients})
and (\ref{q-coefficients}) of Section \ref{sec:eigenvalues}
and in Eq.~(\ref{q-coefficients-extra}) of the Appendix.
\item
We derive the power series for the von Neumann entanglement entropy
for free fermions on a segment of length $L$ of an infinite 
one-dimensional lattice. This expansion is also derived to the tenth 
order in $1/L$, with the coefficients 
listed in Eq.~(\ref{s-coefficients})
of Section \ref{sec:entanglement} and in Eq.~(\ref{s-coefficients-extra}) of the Appendix.
\item
Finally, we check our results for the spectrum and for the entanglement
entropy numerically and find a perfect agreement between our analytic
predictions and numerical data (see Figs.\ \ref{fig-2} and \ref{fig-3}).
This numerical evidence strongly supports the conjecture
(\ref{expansion-1})--(\ref{expansion-2}) and our analytical results.
\end{itemize}

\section{Motivation: free fermions in one dimension}
\label{sec:fermions}

Our study of the spectral determinant (\ref{spectral-determinant}) of the matrix (\ref{matrix-a})
is motivated largely by the problems of full counting statistics (FCS) \cite{levitov:93:96} and entanglement
\cite{calabrese:08,pollmann:10} for
free fermions on the one-dimensional lattice. In the ground state, the single-particle correlation
function is given by the matrix elements (\ref{matrix-a}):
\begin{equation}
\langle \Psi^\dagger_i \Psi_j \rangle = a_{i-j}\, ,
\end{equation}
where $\Psi^\dagger_i$ and $\Psi_j$ are the fermionic creation and annihilation operators
on lattice sites $i$ and $j$, respectively. As shown in Refs.~\onlinecite{chung:01,cheong:04,%
peschel:03,abanov:08:09}, 
the eigenvalues $p_m$ 
of this matrix restricted to some segment of the lattice give the set of {\em single-particle
occupation numbers}, which describe the FCS of fermions on this segment. In particular,
the FCS generating function coincides with $\chi(\kappa,k_F,L)$ defined in eq.~(\ref{spectral-determinant})
and can be written as
\begin{multline}
\chi(\kappa,k_F,L) = 
\left\langle \exp \left(2\pi i \kappa \sum_{i=1}^{L} \Psi^\dagger_i \Psi_i \right) \right\rangle \\ 
= \prod_{m=1}^L \left(1-p_m + p_m e^{2\pi i \kappa} \right)\, .
\end{multline}
In other words, the statistics of the number of particles on the line segment coincides with that
of $L$ levels filled independently with the probabilities $p_m$. 

These probabilities may thus be related to the zeros of the spectral determinant
(\ref{spectral-determinant}). The zeros lie on the negative real axis of $\exp[2\pi i \kappa]$
and therefore may be parametrized as
\begin{equation}
\kappa_m = \frac{1}{2} - i \xi_m
\label{zero-parametrization}
\end{equation}
with real parameters $\xi_m$. The probabilities $p_m$ are, in turn, related to $\xi_m$ as
\begin{equation}
p_m = \frac{1}{1+e^{2\pi\xi_m}}\, .
\label{probabilities}
\end{equation}
The results of the present paper [in particular, equations (\ref{spectrum-equation-1}) and
(\ref{spectrum-equation-2})] describe the spectrum of the probabilities $p_m$
not very close to $0$ or $1$. Such probabilities are relevant for typical fluctuations of the number
of particles around its average. At the same time, our results do not provide
any information about the probabilities $p_m$ very close to $0$ or $1$ (i.e., corresponding to large
$|\xi_m |$): those probabilities would be relevant for atypical fluctuations, e.g., 
for the emptiness formation probability  \cite{korepin:93,shiroishi:01,abanov:03:franchini:05}.

The same spectrum of the probabilities is known to determine
the entanglement (von Neumann) entropy of the
segment with the rest of the lattice \cite{vidal:03,song:11:12,klich:09,calabrese:12}:
\begin{equation}
\mathcal{S}(k_F, L) = \sum_{m=1}^L \left[ -p_m \ln p_m - (1-p_m) \ln (1-p_m) \right]\, . 
\label{entropy-definition}
\end{equation}
The probabilities $p_m$ exponentially close to $0$ or to $1$ give an exponentially small
contribution to the entropy and may be neglected.
This justifies our derivation of the  asymptotic expansion 
(\ref{S-series}) in Section \ref{sec:entanglement} (see also Ref.~\onlinecite{suesstrunk:12} for more details).

\section{Coefficients of the Fisher--Hartwig expansion}
\label{sec:Painleve}

Recurrence relations for the spectral determinant (\ref{spectral-determinant})
were derived from its connection to the theory of discrete Painlev\'e equations
in Ref.~\onlinecite{forrester:03}. Then, in Ref.~\onlinecite{calabrese:10}, the
procedure of extracting the coefficients of the Fisher--Hartwig expansion
was outlined and the first two coefficients computed. We follow the prescription
of Ref.~\onlinecite{calabrese:10} to compute further coefficients of the
Fisher--Hartwig expansion, order by order. 

The recurrence relations read:
\begin{equation}
\frac{\chi(\kappa,k_F,L+1) \, \chi(\kappa,k_F,L-1)}{[\chi(\kappa,k_F,L)]^2}
= 1 - x_L^2\, ,
\label{recurrence-relation-1}
\end{equation}
where $x_L$ obey the relations
\begin{multline}
x_L x_{L-1} - \cos k_F \\
= \frac{1-x_L^2}{2 x_L} \left[(L+1) x_{L+1} + (L-1) x_{L-1} \right] \\
 - \frac{1-x_{L-1}^2}{2 x_{L-1}} \left[L x_{L} + (L-2) x_{L-2} \right]\, .
\label{recurrence-relation-2}
\end{multline}

If we assume the Fisher--Hartwig expansion (\ref{expansion-1})--(\ref{expansion-2}), then,
from the relation (\ref{recurrence-relation-1}),
one finds an expansion for $x_L$ of the form
\begin{equation}
	x_L = \frac{\kappa}{L}(r_L-r_L^{-1})
	+\sum_{n=2}^\infty 
	\frac{Y_n(r_L)}{(r_L-r_L^{-1})^{2n-3}L^n}\, ,
\label{xL-expansion}
\end{equation}
where we have defined 
\begin{equation}
	r_L= (-1)^L \frac{e^{i k_F L}}{(2L\sin k_F)^{2\kappa}}
	\frac{\Gamma(1+\kappa)}{\Gamma(1-\kappa)}
 \label{rL}
\end{equation}
and $Y_n(r_L)$ are Laurent polynomials in $r_L$ with coefficients
depending on $\kappa$ and $k_F$. Each term $Y_n(r_L)$ can be
expressed via the coefficients of the Fisher--Hartwig
expansion $\tilde{F}_{n'}$ with $n'=1,\ldots,n-1$.
By substituting the expansion (\ref{xL-expansion}) into the relation
(\ref{recurrence-relation-2}), 
we can calculate the coefficients 
${\tilde F}_n(\kappa,k_F)$ order by order. Remarkably, the number of conditions
exceeds the number of the coefficients ${\tilde F}_n(\kappa,k_F)$ to be calculated, 
and the fact that the coefficients ${\tilde F}_n(\kappa,k_F)$ satisfy all the 
conditions simultaneously supports the conjecture of the Fisher--Hartwig expansion.
We do not have a proof of this property to all orders, but only observed it in 
calculating the coefficients ${\tilde F}_n(\kappa,k_F)$ to the tenth order.

Note that this procedure is a discrete version of
a similar calculation using Painlev\'e V equation
in the continuous ($k_F\to 0$) limit \cite{mccoy:86,ivanov:13}.
The computations are straightforward, but require tedious
manipulations with polynomials and series. We have performed these
computations using Mathematica software \cite{Wolfram}.

As a result, we find that the coefficients ${\tilde F}_n(\kappa,k_F)$ are polynomials 
in $\kappa$ and $\cot k_F$:
\begin{equation}
{\tilde F}_n(\kappa,k_F) = \sum_{l=0}^{\left\lfloor \frac{n}{2} \right\rfloor}
P_{n,n-2l}(\kappa) \cot^{n-2l} k_F\, ,
\label{F-coefficients}
\end{equation}
where all $P_{nn'}(\kappa)$ are polynomials with real rational coefficients.
We have calculated these polynomials to the tenth order. To avoid lengthy
equations in the main body of the paper, we present below the first six orders,
and report the seventh to tenth orders in Eq.~(\ref{p-coefficients-extra}) of the Appendix:
\begin{align}
P_{11}(\kappa) = {}& 2 \kappa^3\, , \nonumber\\
P_{22}(\kappa) = {}& \frac{5}{2} \kappa^4 \, , \nonumber\\
P_{20}(\kappa) = {}& \frac{4}{3} \kappa^4 + \frac{1}{6} \kappa^2 \, , \nonumber \\
P_{33}(\kappa) = {}& \frac{11}{2} \kappa^5 + \frac{1}{6} \kappa^3 \, , \nonumber \\
P_{31}(\kappa) = {}& \frac{9}{2} \kappa^5 + \frac{1}{2} \kappa^3 \, , \nonumber \\
P_{44}(\kappa) = {}& \frac{63}{4} \kappa^6 + \frac{13}{8} \kappa^4 \, , \nonumber \\
P_{42}(\kappa) = {}& \frac{35}{2} \kappa^6 + \frac{13}{4} \kappa^4 \, , \nonumber \\
P_{40}(\kappa) = {}& \frac{167}{60} \kappa^6 + \frac{25}{24} \kappa^4 + \frac{1}{20} \kappa^2  \, , \label{p-coefficients}\\
P_{55}(\kappa) = {}& \frac{527}{10} \kappa^7 + 12 \kappa^5 + \frac{1}{5} \kappa^3  \, , \nonumber\\
P_{53}(\kappa) = {}& 74 \kappa^7 + \frac{47}{2} \kappa^5 + \frac{1}{2} \kappa^3  \, , \nonumber\\
P_{51}(\kappa) = {}& \frac{45}{2} \kappa^7 + \frac{21}{2} \kappa^5 + \frac{1}{2} \kappa^3  \, , \nonumber\\
P_{66}(\kappa) = {}& \frac{3129}{16} \kappa^8 + \frac{1931}{24} \kappa^6 + \frac{75}{16} \kappa^4  \, , \nonumber\\
P_{64}(\kappa) = {}& \frac{2655}{8} \kappa^8 + \frac{339}{2} \kappa^6 + \frac{47}{4} \kappa^4  \, , \nonumber\\
P_{62}(\kappa) = {}& \frac{2385}{16} \kappa^8 + \frac{781}{8} \kappa^6 + \frac{151}{16} \kappa^4  \, , \nonumber\\
P_{60}(\kappa) = {}& \frac{236}{21} \kappa^8 + \frac{371}{36} \kappa^6 + \frac{43}{24} \kappa^4 
          + \frac{5}{126} \kappa^2 \, , \nonumber
\end{align}
The leading coefficients $P_{nn}(\kappa)$ reproduce $f_n(\kappa)$ found in 
Ref.~\onlinecite{ivanov:13} in the continuous limit $k_F \to 0$.
The coefficients ${\tilde F}_1(\kappa,k_F)$ and ${\tilde F}_2(\kappa,k_F)$ 
have been previously reported in Ref.~\onlinecite{calabrese:10}. The coefficient
${\tilde F}_1(\kappa,k_F)$ was also conjectured in Ref.~\onlinecite{abanov:11}.

We also observe several remarkable properties of these coefficients, of which
we do not have proofs and formulate them as conjectures (to all orders):
\begin{itemize}
\item
The polynomial structure of ${\tilde F}_n(\kappa,k_F)$ persists to all orders,
with the largest degree in $\kappa$ being $n+2$ and the largest degree in 
$\cot k_F$ being $n$.
\item
The coefficients ${\tilde F}_n(\kappa,k_F)$ have the parity of $n$
with respect to $\kappa$ and to $\cot k_F$ (separately).
\item
All the numerical coefficients of these polynomials are real rational numbers.
Most probably, they are all positive.
\item
The terms $\kappa^2$ appear only in the coefficients $P_{2n,0}(\kappa)$.
All the other coefficients $P_{nn'}$  have the lowest terms $\kappa^3$
or $\kappa^4$.
\end{itemize}

\begin{figure}
\centerline{\includegraphics[width=.45\textwidth]{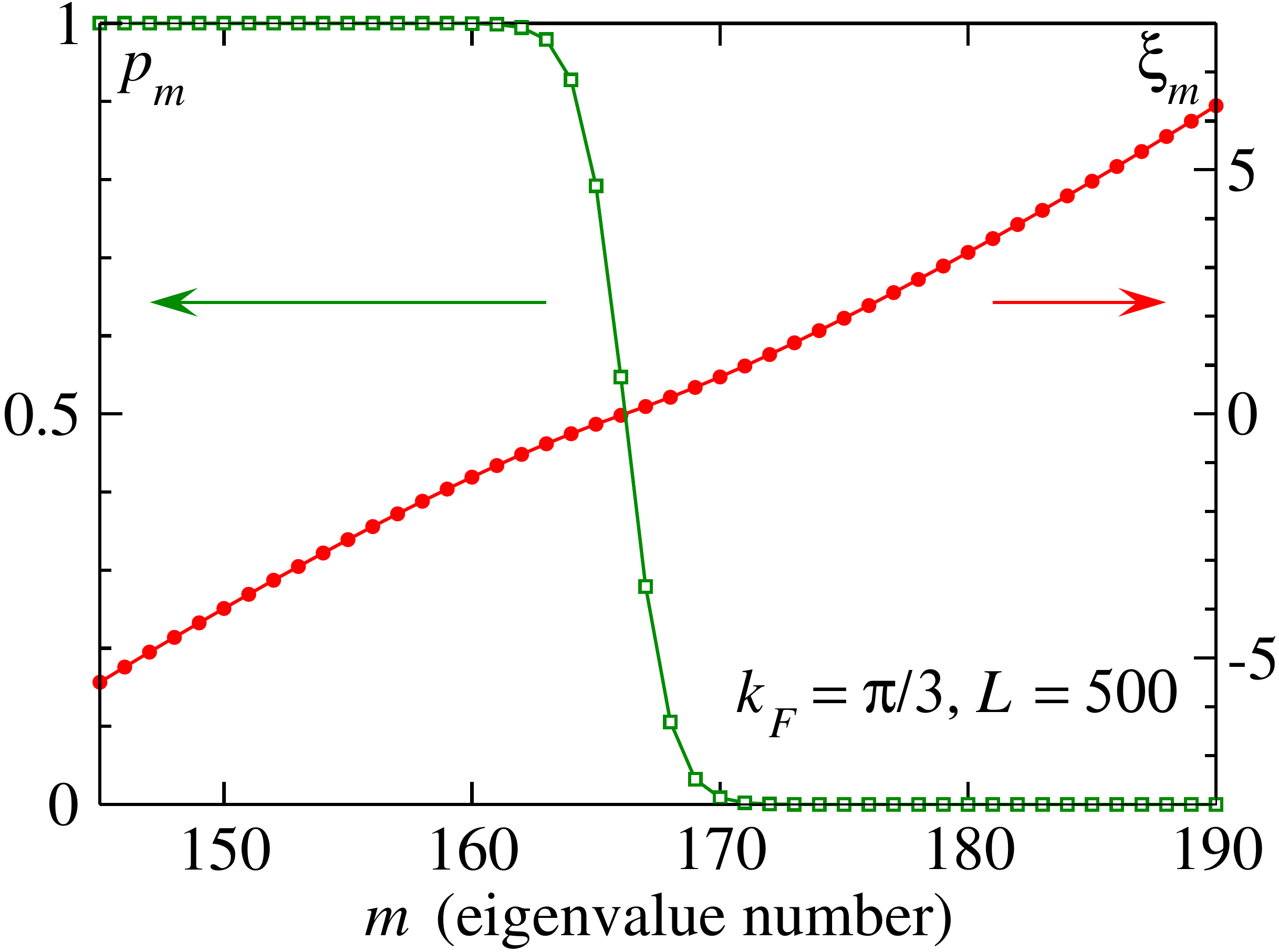}}
\caption{The spectrum of the matrix (\ref{matrix-a}) at 
$k_F=\pi/3$ and $L=500$
in $p_m$ (empty squares, left axis) and $\xi_m$ (solid circles, right axis)
parametrizations.
The two parametrizations are related by Eq.~(\ref{probabilities}).
The roots $p_m$ and $\xi_m$ are enumerated in decreasing and increasing order,
respectively, starting with $m=0$.}
\label{fig-1}
\end{figure}

\begin{figure}
\centerline{\includegraphics[width=.45\textwidth]{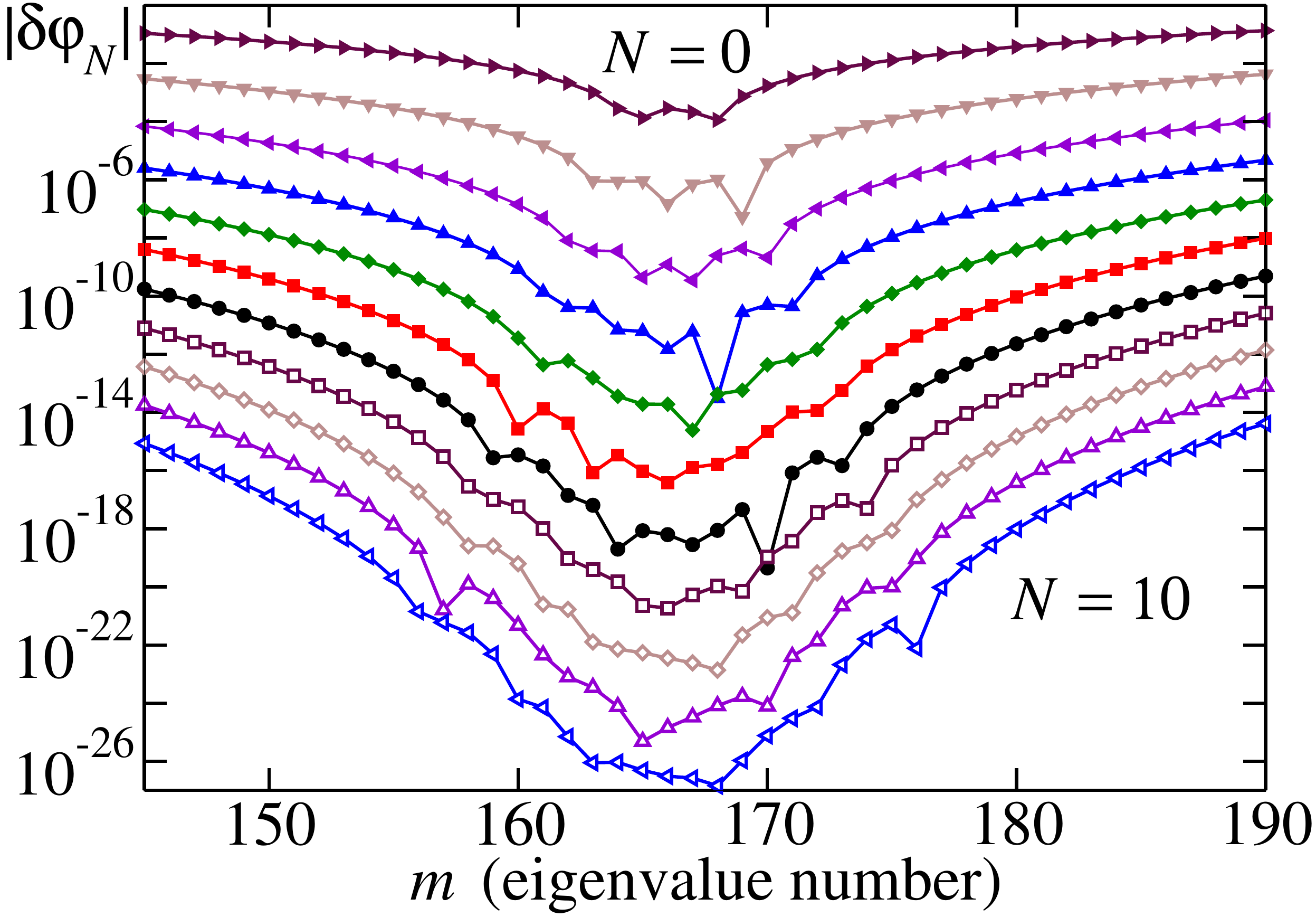}}
\caption{The absolute value of $\delta\varphi_N$, 
the left-hand side of Eq.~(\ref{eigenvalue-remainders}),
as a function of the eigenvalue number for the matrix with 
$k_F=\pi/3$ and $L=500$.
The plots correspond to $N=0$ to $N=10$, from top to bottom. 
The non-smoothness of the plots in the central part of the graph
is not a numerical noise, but related to changes of sign of $\delta\varphi_N$.}
\label{fig-2}
\end{figure}

\begin{figure}
\centerline{\includegraphics[width=.45\textwidth]{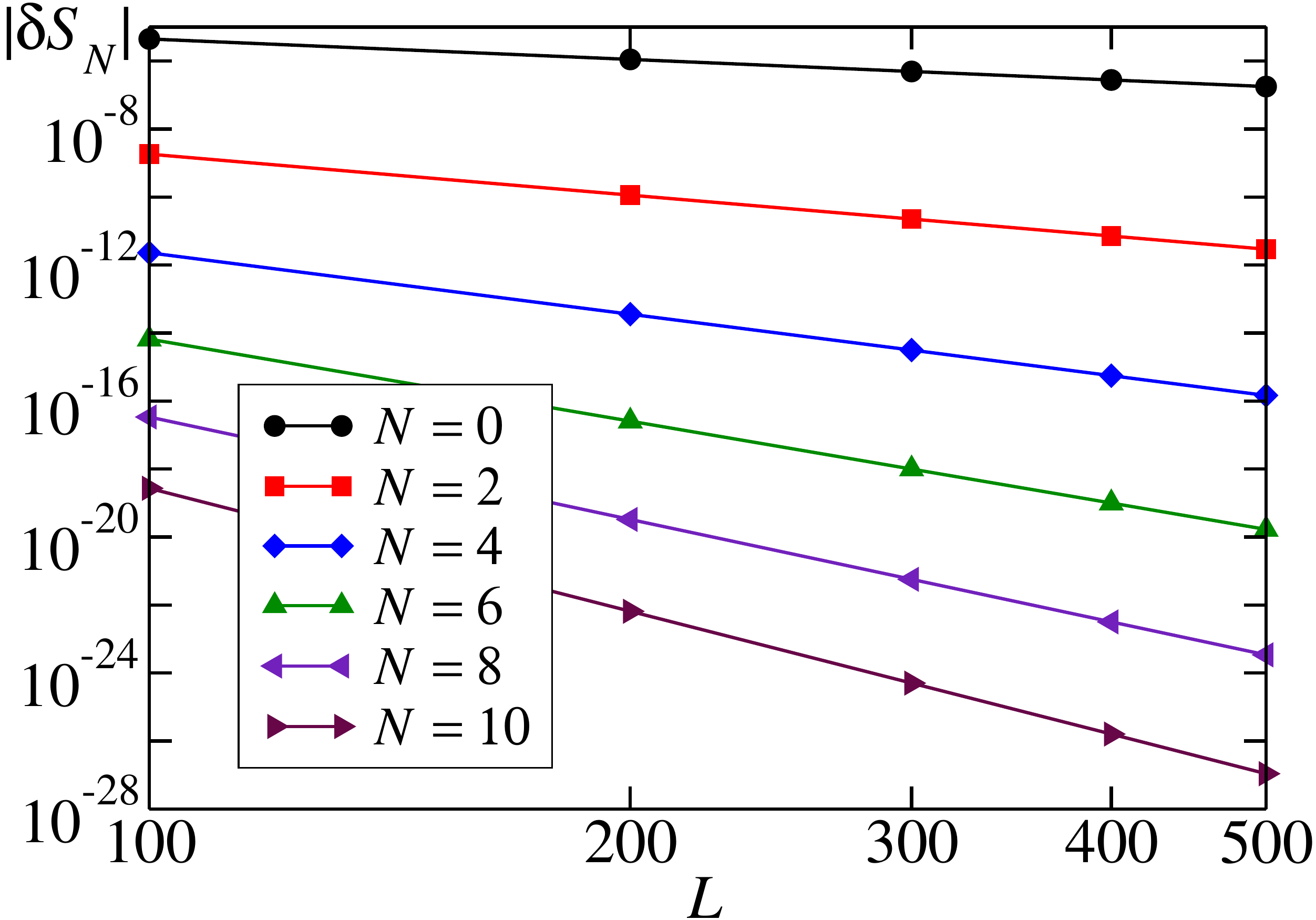}}
\caption{The absolute value of the remainder in the expansion (\ref{S-series})
terminated at order $N$ for the matrix with $k_F=\pi/3$ and several values
$L$ ranging from $100$ to $500$.}
\label{fig-3}
\end{figure}

\section{Eigenvalues of the Toeplitz matrix}
\label{sec:eigenvalues}

If we conjecture that the Fisher--Hartwig expansion 
(\ref{expansion-1})--(\ref{expansion-2}) holds for any $\kappa$
(including those with half-integer real part, where zeros of
$\chi(\kappa,k_F,L)$ accumulate), then we can use it to find
the zeros of $\chi(\kappa,k_F,L)$. 

If we keep only the two leading branches of the Fisher--Hartwig expansion 
($j=0$ and $j=-1$) and only terms up to ${\tilde C}(\kappa)$ in the
exponent (\ref{expansion-2}), then we arrive at the quasiclassical
equation for the zeros
\begin{equation}
\varphi_0(\xi,k_F,L) \approx \pi \left( m + \frac{1}{2} \right)\, ,
\label{spectrum-lowest}
\end{equation}
where we have defined
\begin{equation}
\varphi_0(\xi,k_F,L) = k_F L + 2 \xi \ln (2 L \sin k_F) 
- 2 \arg \Gamma\left(\frac{1}{2}+i\xi \right) \, ,
\label{spectrum-phi}
\end{equation}
$\Gamma()$ denotes the gamma function,
and we use the parametrization (\ref{zero-parametrization}).
This approximation was also derived in Ref.~\onlinecite{suesstrunk:12} 
in the continuous limit and in Ref.~\onlinecite{eisler:13}
using a relation to spheroidal functions \cite{slepian:65:78:83}.

Higher corrections in $1/L$ and higher Fisher--Hartwig branches
may be incorporated in the quasiclassical approximation 
(\ref{spectrum-lowest}) in terms of a power series in $1/L$:
\begin{equation}
	\Phi(\xi_m,k_F,L) = \pi \left( m+\frac{1}{2} \right)\, ,
 \label{spectrum-equation-1}
\end{equation}
where
\begin{equation}
	\Phi(\xi,k_F,L) = \varphi_0(\xi, k_F, L) + \sum_{n=1}^{\infty} \frac{X_n(\xi, k_F)}{L^n}
 \label{spectrum-equation-2}
\end{equation}
[the quantity $\Phi(\xi,k_F,L)$ was introduced earlier in Ref.~\onlinecite{suesstrunk:12},
where the first term in the sum (\ref{spectrum-equation-2}) was computed in the continuous limit].
We do not have a 
proof of this series, but we have checked it
explicitly to the tenth order. 
The coefficients $X_n(\xi, k_F)$ are found to have 
a polynomial form of the same type as ${\tilde F}_n(\kappa,k_F)$: 
\begin{equation}
X_n(\xi, k_F) = \sum_{l=0}^{\left\lfloor \frac{n}{2} \right\rfloor}
Q_{n,n-2l}(\xi) \cot^{n-2l} k_F\, .
\label{X-coefficients}
\end{equation}
For $n$ up to six, the coefficients are:
\begin{align}
Q_{11}(\xi) = {}& 3 \xi^2 - \frac{1}{4}\, , \nonumber\\
Q_{22}(\xi) = {}& - 5\xi^3 +\frac{5}{4}\xi \, , \nonumber\\
Q_{20}(\xi) = {}& - \frac{8}{3} \xi^3 + \frac{5}{6} \xi \, , \nonumber \\
Q_{33}(\xi) = {}& \frac{55}{4} \xi^4 - \frac{57}{8} \xi^2 +\frac{37}{192} \, , \nonumber \\
Q_{31}(\xi) = {}& \frac{45}{4} \xi^4 - \frac{51}{8} \xi^2 + \frac{13}{64} \, , \nonumber \\
Q_{44}(\xi) = {}& -\frac{189}{4} \xi^5 + \frac{85}{2} \xi^3 - \frac{239}{64} \xi \, , \nonumber \\
Q_{42}(\xi) = {}& -\frac{105}{2} \xi^5 + 50 \xi^3 - \frac{155}{32}\xi \, , \nonumber \\
Q_{40}(\xi) = {}& -\frac{167}{20} \xi^5 + \frac{107}{12} \xi^3 - \frac{1019}{960} \xi  \, , \label{q-coefficients}\\
Q_{55}(\xi) = {}& \frac{3689}{20} \xi^6 - \frac{4139}{16} \xi^4 + \frac{15663}{320} \xi^2
         - \frac{1009}{1280} \, , \nonumber\\
Q_{53}(\xi) = {}& 259 \xi^6 - \frac{1515}{4} \xi^4 + \frac{1229}{16} \xi^2
         - \frac{85}{64}  \, , \nonumber\\
Q_{51}(\xi) = {}& \frac{315}{4} \xi^6 - \frac{1965}{16} \xi^4 + \frac{1773}{64} \xi^2
         - \frac{139}{256}  \, , \nonumber\\
Q_{66}(\xi) = {}& -\frac{3129}{4} \xi^7 + \frac{25459}{16} \xi^5 - \frac{102949}{192} \xi^3
         + \frac{7245}{256} \xi  \, , \nonumber\\
Q_{64}(\xi) = {}& -\frac{2655}{2} \xi^7 + \frac{22323}{8} \xi^5 - \frac{31745}{32} \xi^3
         + \frac{7093}{128} \xi  \, , \nonumber\\
Q_{62}(\xi) = {}& -\frac{2385}{4} \xi^7 + \frac{20979}{16} \xi^5 - \frac{32095}{64} \xi^3
         + \frac{7789}{256} \xi  \, , \nonumber\\
Q_{60}(\xi) = {}& -\frac{944}{21} \xi^7 + \frac{1279}{12} \xi^5 - \frac{1661}{36} \xi^3 
          + \frac{13261}{4032} \xi \, , \nonumber
\end{align}
and the seventh to tenth orders are presented in Eq.~(\ref{q-coefficients-extra}) of the Appendix.
From observing the polynomial structure of the coefficients, we conjecture that
the equation (\ref{spectrum-equation-1})--(\ref{spectrum-equation-2}) 
holds asymptotically at $L \to \infty$.
Namely, we conjecture that 
at any fixed $k_F$, at any order $N$, and for any $\Xi$, the remainder 
can be estimated as
\begin{equation}
\varphi_0(\xi_m, k_F, L) + \sum_{n=1}^{N} \frac{X_n(\xi_m, k_F)}{L^n} - 
\pi \left( m+\frac{1}{2} \right) = o(L^{-N})
\label{eigenvalue-remainders}
\end{equation}
uniformly for all $\xi_m$ within the window
\begin{equation}
|\xi_m| < \Xi\, .
\end{equation}
This expansion extends the lowest-order approximation to the spectrum obtained
earlier in  Refs.~\onlinecite{suesstrunk:12,eisler:13,slepian:65:78:83}. 
It is also in agreement with findings of Ref.~\onlinecite{peschel:04:09}.
For practical purposes, cutting off the series at a finite $N$ provides a good
approximation, if $k_F L \gg c_N \max(1,|\xi|)$ (with some coefficients $c_N$ which depend
on the rate of growth of the numerical coefficients in eqs.~(\ref{q-coefficients}) with
the order and which we do not study here).

Finally remark that the sum of all the probabilities (\ref{probabilities})
must give the total average number of particles, i.e.,
\begin{equation}
\sum_m \frac{1}{1+e^{2\pi\xi_m}} = \frac{k_F L}{\pi}\, .
\label{normalization}
\end{equation}
This condition allows us identify the integer index $m$ in Eq.~(\ref{spectrum-equation-1})
with the sequential number of the root $\xi_m$ in the increasing order starting with $m=0$.

\section{Von Neumann entanglement for one-dimensional free fermions}
\label{sec:entanglement}

Similarly to the roots of $\chi(\kappa,k_F,L)$, we can calculate the expansion
for the von Neumann entanglement entropy (\ref{entropy-definition}). The whole discussion
of Ref.~\onlinecite{suesstrunk:12} applies to the lattice case, with the
only difference in the coefficients ${\tilde F}_n(\kappa,k_F)$. In particular,
if the conjecture about the polynomial structure of the coefficients
${\tilde F}_n(\kappa,k_F)$ (with respect to $\kappa$) is valid, then
the statement of Ref.~\onlinecite{suesstrunk:12} about the power-law expansion
of the entropy (without oscillating terms) extends directly to the lattice case.
A straightforward calculation along the lines of Ref.~\onlinecite{suesstrunk:12}
produces the expansion
\begin{equation}
 \mathcal{S}(k_F, L)=\frac{1}{3}\ln(2 L \sin k_F)+ \Upsilon + \sum_{n=1}^{\infty} {\tilde s}_{2n}(k_F) L^{-2n}\, ,
\label{S-series}
\end{equation}
where
\begin{multline}
\Upsilon = -\frac{2}{\pi} \int_{-\infty}^{+\infty} d\xi \, \re \psi\left(\frac{1}{2}+i\xi\right) \\
\times
\left(\ln\left[2\cosh(\pi \xi)\right] - \pi \xi \tanh[\pi \xi] \right) \\
= 0.49501790813513705018901197430445590 \ldots
\label{Upsilon}
\end{multline}
is a numerical constant
[$\psi()$ is the digamma function: the logarithmic derivative of $\Gamma()$]
and ${\tilde s}_{2n}(k_F)$ are polynomials. A calculation to the tenth order gives:
\begin{align}
{\tilde s}_{2}(k_F) = {}& -\frac{1}{12}\cot^2 k_F - \frac{1}{60}\, , \nonumber\\
{\tilde s}_{4}(k_F) = {}& -\frac{31}{96}\cot^4 k_F - \frac{5}{16} \cot^2 k_F - \frac{47}{1120} \, , 
\label{s-coefficients} \\
{\tilde s}_{6}(k_F) = {}& -\frac{7057}{1440}\cot^6 k_F - \frac{247}{32} \cot^4 k_F 
- \frac{301}{96} \cot^2 k_F \nonumber\\
& - \frac{403}{2016} \, , \nonumber
\end{align}
and the coefficients ${\tilde s}_{8}(k_F)$ and ${\tilde s}_{10}(k_F)$ are presented
in Eq.~(\ref{s-coefficients-extra}) of the Appendix.
The logarithmic term is well-known from the conformal-field-theory considerations \cite{vidal:03},
the $\Upsilon$ term was computed earlier in Ref.~\onlinecite{jin:04},
the ${\tilde s}_{2}(k_F)$ is known from Refs.~\onlinecite{calabrese:10,calabrese:11},
and the leading orders in ${\tilde s}_{2}(k_F)$, ${\tilde s}_{4}(k_F)$, and
${\tilde s}_{6}(k_F)$ were reported in the continuous limit in Ref.~\onlinecite{suesstrunk:12}.

An alternative way of calculating the expansion (\ref{S-series}) is by
using the expansion (\ref{spectrum-equation-2})--(\ref{q-coefficients})
for $\Phi(\xi,k_F,L)$. Indeed, it follows from the discussion of 
Ref.~\onlinecite{suesstrunk:12} that the von Neumann entropy can
be calculated as
\begin{equation}
{\cal S} (k_F,L)
= \int_{-\infty}^{+\infty} \frac{\pi\xi\, d\xi}{\cosh^2(\pi\xi)} \Phi(\xi,k_F,L)\, ,
\end{equation}
which immediately gives the coefficients ${\tilde s}_{2n}$ once the
coefficients $X_{2n}$ are known.

\section{Numerical illustration}
\label{sec:numerics}

Since the results presented in this paper are not proven in a rigorous way, but
rely on the conjecture of the Fisher--Hartwig expansion and of the structure
of the expansion coefficients, we find it helpful to check them against numerical
data. For such a test, we take the wave vector $k_F=\pi/3$ (which corresponds
to the filling fraction $1/3$). We perform an exact diagonalization of large
matrices (\ref{matrix-a}) with the use of LAPACK library \cite{Lapack} compiled to work
with 128-bit-precision floating-point numbers, together with the quadmath
C library.

An example of the spectrum found in this way is plotted in Fig.~\ref{fig-1}
in both $p_m$ and $\xi_m$ parametrizations. Note that we enumerate $p_n$ in decreasing
and $\xi_m$ in increasing order, starting with $m=0$. The normalization condition
(\ref{normalization}) guarantees that $p_m$ crosses over from $1$ to $0$ at
$n \approx k_F L / \pi$.

Next, we test the expansion for the eigenvalue equation 
(\ref{spectrum-equation-1})--(\ref{spectrum-equation-2}).
We denote by $\delta\varphi_N$ the left-hand side of Eq.~(\ref{eigenvalue-remainders}).
In Fig.~\ref{fig-2}, we plot $|\delta\varphi_N|$ as a function of $n$ (the number of eigenvalue)
for the same example of $k_F=\pi/3$ and $L=500$, using the coefficients $X_n(\xi,k_F)$
listed in Eqs. (\ref{X-coefficients}) and (\ref{q-coefficients}). From the exponential
decay of $\delta\varphi_N$ with increasing $N$, we see that the coefficients are correct.

Finally, we also test the expansion of the von Neumann entanglement entropy (\ref{S-series}).
Denote by $\delta{\cal S}_N$ the difference between the exact
value of the entropy ${\cal S}(k_F, L)$ and the right-hand side of Eq.~(\ref{S-series})
with the sum truncated at $N$ terms. The absolute value of $\delta {\cal S}_N$ 
is plotted in Fig.~\ref{fig-3} for the case of $k_F=\pi/3$ and several different
values of $L$. From the exponential decay of $\delta{\cal S}_N$ with increasing $N$,
we see that the coefficients are correct.

Our numerical test confirms the analytical expressions for the matrix eigenvalues
and the entanglement entropy. Note that the higher-order coefficients (starting with
the fourth order) contain contributions from several Fisher--Hartwig branches.
We are therefore confident that the original conjecture about the validity of the
Fisher--Hartwig expansion for all values of $\kappa$ holds.

\null

\section{Conclusion}
\label{sec:conclusion}

In this paper, we employ
the conjectured Fisher--Hartwig expansion
for the sine-kernel Toeplitz matrix for calculating the expansion coefficients.
Furthermore, we translate those results into finite-size corrections for the von Neumann
entanglement entropy and for the quasiclassical-type equation on the spectrum. 

We hope that these results will serve as a useful reference for future studies of Toeplitz
determinants. Besides, they provide a strong support to the conjectured Fisher--Hartwig
expansion in the periodic form (\ref{expansion-1})--(\ref{expansion-2}). While it
is not proven, our calculation to the tenth order leaves
no doubt in its validity in the case of the sine kernel (\ref{matrix-a}).

Our study also outlines further challenges in the theory of Toeplitz determinants.
In particular, we find the following open questions deserving 
future consideration:
\begin{itemize}
\item
Proving the Fisher--Hartwig expansion (\ref{expansion-1})--(\ref{expansion-2}),
at least in the case of the sine kernel (\ref{matrix-a}) and possibly in a
general case of a Toeplitz matrix with Fisher--Hartwig singularities. 
Furthermore,
one could attempt an even more general extension to determinants of the
form $\det(1+AB)$ where $A$ and $B$ are local operators in the coordinate
and momentum space, respectively: 
the leading Fisher--Hartwig asymptotics for
such determinants was
studied in 
Ref.~\onlinecite{protopopov:12} in the context of nonequilibrium bosonization.
\item
Exploring the decomposition into Fisher--Hartwig branches (\ref{expansion-1}):
can one extend such a decomposition to finite values of $L$
[with $\chi_*(\kappa, k_F, L)$ defined at any finite $L$,
and not as a formal asymptotic series (\ref{expansion-2})]? 
\item
In the case of the sine kernel (\ref{matrix-a}), proving the polynomial structure
of the coefficients ${\tilde F}_n (\kappa,k_F)$ to all orders, as well as 
the properties of those polynomials conjectured in Section \ref{sec:Painleve}.
Such a proof would be most probably related to 
the integrability of the corresponding Painlev\'e equations.
\end{itemize}

\begin{acknowledgments}
D.I.\ thanks 
I.~Protopopov, A.~Mirlin, 
P.~Schmitteckert and S.~Winitzki 
for discussions and the Simons Center for Geometry  and Physics for hospitality. 
The work of AGA was supported by the NSF under Grant No.\ DMR-1206790. 
\end{acknowledgments}


\section*{Appendix: Seventh to tenth orders}

We present here the coefficients of the expansions (\ref{F-coefficients}), (\ref{X-coefficients}),
and (\ref{S-series}) in orders seven to ten.

\begin{align}
P_{77}(\kappa) = {}& \frac{175045}{224} \kappa ^9 +\frac{8263}{16} \kappa^7 \nonumber\\
& +\frac{2155}{32} \kappa ^5 + \frac{45}{56} \kappa ^3 \, , \nonumber\\
P_{75}(\kappa) = {}& \frac{49755}{32} \kappa ^9 +\frac{19175}{16} \kappa ^7 \nonumber\\
& +\frac{5699 }{32}\kappa ^5+\frac{19}{8} \kappa ^3 \, , \nonumber\\
P_{73}(\kappa) = {}& \frac{88735}{96} \kappa ^9 +\frac{13585}{16} \kappa ^7 \nonumber\\
& +\frac{4909}{32} \kappa ^5+\frac{59}{24} \kappa ^3 \, , \nonumber \\
P_{71}(\kappa) = {}& \frac{4765}{32} \kappa ^9 +\frac{2721}{16} \kappa ^7 \nonumber\\
& +\frac{1317}{32} \kappa ^5 +\frac{9}{8} \kappa ^3 \, , \nonumber \\
P_{88}(\kappa) = {}& \frac{422565}{128}\kappa ^{10} + \frac{414915}{128} \kappa^8 \nonumber\\
& +\frac{98575}{128} \kappa ^6 + \frac{4183}{128} \kappa^4 \, , \nonumber \\
P_{86}(\kappa) = {}& \frac{723283}{96} \kappa ^{10}+ \frac{265959}{32} \kappa^8 \nonumber\\
& +\frac{70011}{32} \kappa ^6 +\frac{9761 }{96} \kappa ^4 \, , \nonumber \\
P_{84}(\kappa) = {}& \frac{356807}{64} \kappa ^{10}+\frac{448929}{64} \kappa^8 \label{p-coefficients-extra}\\
& +\frac{135365}{64} \kappa ^6 +\frac{7165}{64} \kappa ^4 \, , \nonumber \\
P_{82}(\kappa) = {}& \frac{44825}{32} \kappa ^{10} +\frac{65967}{32} \kappa^8 \nonumber\\
& +\frac{23907}{32} \kappa ^6 +\frac{1587 }{32}\kappa ^4 \, , \nonumber\\
P_{80}(\kappa) = {}& \frac{353777}{5760} \kappa ^{10}+\frac{42761}{384} \kappa^8 \nonumber\\
& +\frac{102449}{1920} \kappa ^6 +\frac{6527}{1152} \kappa^4 +\frac{7}{120} \kappa ^2 \, , \nonumber\\
P_{99}(\kappa) = {}& \frac{1398251}{96} \kappa ^{11} +\frac{180872}{9} \kappa^9 \nonumber\\
& +\frac{246729}{32} \kappa ^7 +\frac{35483}{48} \kappa^5 +7 \kappa ^3 \, , \nonumber\\
P_{97}(\kappa) = {}& \frac{149997}{4} \kappa ^{11} + \frac{907719}{16} \kappa^9 \nonumber\\
& +\frac{189241}{8} \kappa ^7+\frac{39147}{16} \kappa^5+\frac{99}{4} \kappa ^3 \, , \nonumber\\
P_{95}(\kappa) = {}& \frac{2656689 }{80}\kappa ^{11}+\frac{891143}{16} \kappa^9 \nonumber\\
& +\frac{2059487}{80} \kappa ^7 + \frac{47061}{16} \kappa^5 + \frac{651}{20} \kappa ^3 \, , \nonumber\\
P_{93}(\kappa) = {}& 11414 \kappa ^{11}+\frac{1035763}{48} \kappa ^9 \nonumber\\
& +\frac{90937}{8} \kappa ^7 +\frac{23965}{16} \kappa ^5 +\frac{229}{12} \kappa^3  \, , \nonumber\\
P_{91}(\kappa) = {}& \frac{36597 }{32}\kappa ^{11} +\frac{40129}{16} \kappa^9 \nonumber\\
& +\frac{50457}{32} \kappa ^7+\frac{2085}{8} \kappa^5 +\frac{19 }{4}\kappa ^3  \, , \nonumber\\
P_{10,10}(\kappa) = {}& \frac{266149}{4} \kappa ^{12} +\frac{1979539}{16} \kappa^{10} \nonumber\\
& +\frac{2264875}{32} \kappa ^8+\frac{198967}{16} \kappa^6 +\frac{6795}{16} \kappa ^4 \, , \nonumber\\
P_{10,8}(\kappa) = {}&  \frac{381511 }{2}\kappa ^{12}+\frac{24475367 }{64}\kappa^{10} \nonumber\\
& +\frac{15007249}{64} \kappa ^8+\frac{2812389}{64} \kappa^6+\frac{101805}{64} \kappa ^4 \, , \nonumber\\
P_{10,6}(\kappa) = {}& \frac{786877}{4} \kappa ^{12} +\frac{6860081}{16} \kappa^{10} \nonumber\\
& +\frac{4567499}{16} \kappa ^8+\frac{927265}{16} \kappa^6 +2260 \kappa ^4  \, , \nonumber\\
P_{10,4}(\kappa) = {}& \frac{344253}{4} \kappa ^{12} +\frac{6602861}{32} \kappa^{10} \nonumber\\
& +\frac{4858935}{32} \kappa ^8 +\frac{1094911}{32} \kappa^6+\frac{47327}{32} \kappa ^4 \, , \nonumber\\
P_{10,2}(\kappa) = {}& \frac{27937 }{2}\kappa ^{12}+\frac{300145 }{8}\kappa^{10} \nonumber\\
& +\frac{1002999}{32} \kappa ^8+\frac{65335}{8} \kappa^6+\frac{6677}{16} \kappa ^4 \, , \nonumber\\         
P_{10,0}(\kappa) = {}& \frac{264031 }{660}\kappa ^{12} +\frac{1194617}{960} \kappa^{10}
+\frac{397029}{320} \kappa ^8 \nonumber\\
& +\frac{78175 }{192}\kappa^6 +\frac{9561 }{320}\kappa ^4 +\frac{3}{22} \kappa ^2  \, . \nonumber        
\end{align}

\begin{align}
Q_{77}(\xi) = {}& \frac{1575405}{448} \xi^8 - \frac{630205}{64}\xi ^6
  +\frac{2703025}{512} \xi^4 \nonumber\\
  & -\frac{4328955}{7168} \xi^2+\frac{822221}{114688}\, , \nonumber\\
Q_{75}(\xi) = {}& \frac{447795}{64} \xi ^8 - \frac{1288133}{64} \xi ^6
  +\frac{5770025}{512} \xi ^4 \nonumber\\
  & -\frac{1383957}{1024} \xi ^2 + \frac{274707}{16384}\, , \nonumber\\
Q_{73}(\xi) = {}& \frac{266205}{64} \xi ^8 - \frac{792827}{64} \xi ^6
  +\frac{3753895}{512} \xi ^4 \nonumber\\
  & -\frac{958667}{1024} \xi ^2 +\frac{202173}{16384}\, , \nonumber \\
Q_{71}(\xi) = {}& \frac{42885}{64} \xi ^8-\frac{134211}{64} \xi ^6
+\frac{687615}{512} \xi ^4 \nonumber \\
 & -\frac{193011}{1024} \xi ^2 +\frac{44965}{16384} \, , \nonumber \\
Q_{88}(\xi) = {}& -\frac{2112825}{128} \xi ^9+\frac{1957401}{32} \xi ^7
-\frac{49541793}{1024} \xi ^5 \nonumber \\
 & +\frac{10152729}{1024} \xi ^3-\frac{12653165}{32768} \xi  \, , \nonumber \\
Q_{86}(\xi) = {}& -\frac{3616415}{96} \xi ^9+\frac{571261}{4}\xi ^7
-\frac{29986677}{256} \xi ^5 \nonumber \\
 & +\frac{19195649}{768} \xi ^3-\frac{8290601}{8192}\xi  \, , \nonumber \\
Q_{84}(\xi) = {}& -\frac{1784035}{64} \xi ^9+\frac{1739145}{16} \xi ^7
-\frac{47722731}{512} \xi ^5 \nonumber\\
 & +\frac{10713015}{512} \xi ^3-\frac{14593215}{16384} \xi \, , \nonumber \\
Q_{82}(\xi) = {}& -\frac{224125}{32} \xi ^9+\frac{56685}{2}\xi ^7
-\frac{6594861 }{256}\xi ^5 \nonumber\\
 & +\frac{1585615}{256} \xi ^3-\frac{2318257}{8192} \xi  \, , \label{q-coefficients-extra}\\
Q_{80}(\xi) = {}& -\frac{353777}{1152} \xi ^9+\frac{126127}{96} \xi ^7
-\frac{20024543}{15360} \xi ^5 \nonumber \\
 & +\frac{3225689}{9216} \xi ^3-\frac{8925299}{491520} \xi \, , \nonumber\\
Q_{99}(\xi) = {}&  \frac{15380761 }{192}\xi ^{10}-\frac{97519635}{256} \xi ^8 \nonumber \\
 &+\frac{646571513 }{1536}\xi ^6 -\frac{848130205}{6144} \xi ^4 \nonumber\\
 &+\frac{191133511}{16384} \xi^2
 -\frac{66947335}{589824}\, , \nonumber\\
Q_{97}(\xi) = {}& \frac{1649967}{8} \xi ^{10}-\frac{32008545 }{32}\xi ^8 \nonumber\\
 & +\frac{72975679}{64} \xi ^6-\frac{99152045}{256} \xi^4 \nonumber\\
 & +\frac{69407139}{2048} \xi ^2-\frac{2787573}{8192} \, , \nonumber\\
Q_{95}(\xi) = {}&  \frac{29223579}{160} \xi ^{10} -\frac{116094807}{128} \xi ^8 \nonumber\\
 & +\frac{1373306011}{1280} \xi ^6 -\frac{389409139}{1024} \xi^4 \nonumber\\
 & +\frac{1422641679}{40960} \xi ^2 -\frac{11885411}{32768}\, , \nonumber\\
Q_{93}(\xi) = {}&  62777 \xi ^{10}-\frac{5137695 }{16}\xi ^8 \nonumber\\
 & +\frac{3181941 }{8}\xi ^6 -\frac{19048415 }{128}\xi ^4 \nonumber\\
 & +\frac{3680047}{256} \xi^2-\frac{1944337}{12288} \, , \nonumber\\
Q_{91}(\xi) = {}&  \frac{402567}{64} \xi ^{10}-\frac{8573175}{256} \xi ^8 \nonumber\\
 & +\frac{22598247}{512} \xi ^6-\frac{36466515}{2048} \xi^4 \nonumber\\
 & +\frac{30587211}{16384} \xi ^2 -\frac{1461187}{65536} \, , \nonumber\\
Q_{10,10}(\xi) = {}& -\frac{798447}{2} \xi ^{11} +\frac{304191195}{128} \xi ^9 \nonumber\\
 & -\frac{450699243}{128} \xi ^7+\frac{1754196861}{1024} \xi^5 \nonumber\\
 & -\frac{531991659}{2048} \xi ^3+\frac{271418723}{32768} \xi \, , \nonumber\\
 \nonumber\\[-11pt]
Q_{10,8}(\xi) = {}&  -1144533 \xi ^{11}+\frac{443665365}{64} \xi ^9 \nonumber\\
& -\frac{337710603 }{32}\xi ^7+\frac{2711784825}{512} \xi^5 \nonumber\\
& -\frac{212156437}{256} \xi ^3+\frac{445785857 }{16384}\xi  \, , \nonumber\\
Q_{10,6}(\xi) = {}& -\frac{2360631 }{2}\xi ^{11}+\frac{466926845}{64} \xi ^9 \nonumber\\
& -\frac{733633101}{64} \xi ^7+\frac{3055270995}{512} \xi^5 \nonumber\\
& -\frac{992663763}{1024} \xi ^3+\frac{540165541}{16384} \xi   \, , \nonumber\\
Q_{10,4}(\xi) = {}& -\frac{1032759}{2} \xi ^{11} +3271345 \xi ^9 \nonumber\\
& -\frac{170842821 }{32}\xi ^7+\frac{371988813}{128} \xi^5 \nonumber\\
& -\frac{63318051}{128} \xi ^3+\frac{36028627}{2048}  \xi \, , \nonumber\\
Q_{10,2}(\xi) = {}& -83811 \xi ^{11}+\frac{70127475}{128} \xi ^9 \nonumber\\
& -\frac{120088947}{128} \xi ^7+\frac{554032437 }{1024}\xi^5 \nonumber\\
& -\frac{200707063}{2048} \xi ^3+\frac{121435547}{32768} \xi  \, , \nonumber\\         
Q_{10,0}(\xi) = {}&  -\frac{264031}{110} \xi ^{11}+\frac{3147805}{192} \xi ^9 \nonumber\\
& -\frac{2402497}{80} \xi ^7+\frac{9642191}{512} \xi^5 \nonumber\\
& -\frac{4804861}{1280} \xi ^3+\frac{85042939}{540672} \xi  \, . \nonumber       
\end{align}

\begin{align}
{\tilde s}_{8}(k_F) = {}& -\frac{2578531}{15360} \cot^8 k_F - \frac{1398467}{3840} \cot^6 k_F \nonumber\\
  & - \frac{388885}{1536} \cot^4 k_F - \frac{45031}{768} \cot^2 k_F - \frac{77197}{33792} \, , \nonumber\\
{\tilde s}_{10}(k_F) = {}& -\frac{110519141}{10752} \cot^{10} k_F -\frac{508262583}{17920} \cot^8 k_F \nonumber\\
  & -\frac{35797497}{1280} \cot^6 k_F -\frac{14834081}{1280} \cot^4 k_F  \nonumber\\
  & - \frac{4494921}{2560} \cot^2 k_F -\frac{83689503}{1830400}\, . 
  \label{s-coefficients-extra}
\end{align}




\begin{thebibliography}{99}
\bibitem{note-notations}
  Our parameter $\kappa$ coincides with that in Refs.~\onlinecite{suesstrunk:12,ivanov:13} and
  is related to the counting fields $\lambda$ in Refs.~\onlinecite{levitov:93:96,abanov:08:09,abanov:11}
  and $\eta$ in Ref.~\onlinecite{cheianov:04} by $2\pi\kappa=\lambda=\eta$. It is also related to the
  parameter $\beta_\lambda$ in Ref.~\onlinecite{calabrese:10} as $\kappa=-\beta_\lambda$. Our
  definition of ${\tilde F_n}(\kappa,k_F)$ differs from  $F_n(\kappa,k_F)$ in Ref.~\onlinecite{abanov:11}
  by a factor of $i^n$, in order to make it a real function. Our definition of ${\tilde C}(\kappa)$ differs
  from $C(\kappa)$ in Ref.~\onlinecite{ivanov:13} by $2\kappa^2 \ln2$, in order to simplify formulas.
\end{thebibliography}
\end{document}